\documentclass[aps,prb,twocolumn,superscriptaddress,showpacs,floatfix]{revtex4}
\usepackage{graphicx,epsfig,amsmath}

\newcommand{\bohrm}{\mbox{$\mu_{B}$}}
\newcommand{\kboltz}{\mbox{$k_{B}$}}

\newcommand{\etal}{\textit{et al.}}

\newcommand{\AlxGaAs}[2]{\mbox{$\text{Al}_{#1}\text{Ga}_{#2}\text{As}$}}
\newcommand{\Vds}{\mbox{$\text{V}_{\text{ds}}$}}
\newcommand{\Vg}{\mbox{$\text{V}_{\text{g}}$}}
\newcommand{\VgZero}{\mbox{$\text{V}_{\text{g,0}}$}}
\newcommand{\plaindidv}{\mbox{dI/d\Vds}}

\newcommand{\ZeemanE}{\mbox{$|g|\bohrm B$}}

\newcommand{\TKondo}{\mbox{$\text{T}_{\text{K}}$}}
\newcommand{\TKondoZero}{\mbox{$\text{T}_{\text{K,0}}$}}
\newcommand{\TKondoC}{\mbox{$\text{T}_{\text{K,C}}$}}
\newcommand{\THWHM}{\mbox{$\text{T}_{\text{HWHM}}$}}
\newcommand{\DeltaK}{\mbox{$\Delta_{\text{K}}$}}
\newcommand{\DeltaKZero}{\mbox{$\Delta_{\text{K,0}}$}}
\newcommand{\Bc}{\mbox{$B_{\text{C}}$}}

\newcommand{\startsubfig}[2]{Figure~\ref{fig:#1}(#2)}
\newcommand{\subfig}[2]{Fig.~\ref{fig:#1}(#2)}
\newcommand{\allfig}[1]{Fig.~\ref{fig:#1}}
\newcommand{\refeqn}[1]{Equation (\ref{eq:#1})}

\begin{document}

%\preprint{}

% The title
\title{Kondo-temperature dependence of the Kondo splitting in a single-electron transistor}

% The authors
\author{S. Amasha}
	\affiliation{Department of Physics, Massachusetts Institute of Technology, Cambridge, Massachusetts 02139}

\author{I. J. Gelfand}
	\affiliation{Division of Engineering and Applied Sciences, Harvard University, Cambridge, Massachusetts 02138}

\author{M. A. Kastner} 
	\email{mkastner@mit.edu}
	\affiliation{Department of Physics, Massachusetts Institute of Technology, Cambridge, Massachusetts 02139}
	
\author{A. Kogan}
	\altaffiliation[Current address: ]{Department of Physics, University of Cincinnati, Cincinnati, OH 45221-0011}	
	\affiliation{Department of Physics, Massachusetts Institute of Technology, Cambridge, Massachusetts 02139}

% date
%\date{\today}

% The abstract
\begin{abstract}
	
A Kondo peak in the differential conductance of a single-electron transistor is measured as a function of both magnetic field and the Kondo temperature. We observe that the Kondo splitting decreases logarithmically with Kondo temperature and that there exists a critical magnetic field \Bc~ below which the Kondo peak does not split, in qualitative agreement with theory. However, we find that the magnitude of the prefactor of the logarithm is larger than predicted and is independent of $B$, in contradiction with theory. Our measurements also suggest that the value of \Bc~ is smaller than predicted.
	 
\end{abstract}

% The PACS:

% make the title
\maketitle

% begin the article:	

	The many-electron Kondo state is formed when conduction electrons screen a magnetic impurity. An important tool for studying the Kondo effect is the single-electron transistor (SET)\cite{glazman88:JETPlett, ng88:OnsiteRepulsion, goldhaber98:NatureKondo, cronenwett98:ScienceKondo} which consists of a confined droplet of electrons, called an artificial atom or a quantum dot, coupled by tunnel barriers to two conducting leads, called the source and the drain. When the quantum dot contains a net spin the conduction electrons in the leads screen the spin on the dot and enhance the conductance of the SET. By using the gate electrodes that define the artificial atom to tune its parameters, Goldhaber-Gordon \etal\cite{goldhaber98:PRLKondo} have tested the predictions of renormalization group calculations and scaling theory and have shown that the equilibrium properties of the SET are quantitatively described by the Anderson Hamiltonian.
	
	SETs also provide the unique possibility of exploring non-equilibrium Kondo phenomena by applying a DC voltage \Vds\space between the drain and the source. A sensitive test\cite{meir93:nonequilAnderson, costi00:KondoSplitting, moore00:KondoSplitting,rosch03:NonequilibriumTransport,rosch03:SpecFuncInHighB} of theories of non-equilibrium Kondo physics is the measurement of the splitting in a magnetic field $B$ of the spin-1/2 Kondo peak in differential conductance as a function of \Vds. Calculations by Meir \etal\cite{meir93:nonequilAnderson} predict that in a magnetic field the Kondo peaks occur at $e\Vds= \pm\Delta$, where $\Delta = \ZeemanE$ is the Zeeman energy for spin splitting and $\bohrm= 58$  \mbox{$\mu$}eV/T is the Bohr magneton. We define $\DeltaK/e$ as half the separation in \Vds\space between the two peaks, so Meir \etal~ predict $\DeltaK = \Delta$. Early measurements by Cronenwett \etal\cite{cronenwett98:ScienceKondo} agree with this prediction. However calculations of spectral functions predict new features in the Kondo splitting. Costi\cite{costi00:KondoSplitting} predicts that the screening of the spin on the quantum dot by the conduction electrons should cause the Kondo peak to split only above a critical magnetic field \Bc, which depends on the Kondo temperature \TKondo, the energy scale that describes the strength of the screening. Moore and Wen\cite{moore00:KondoSplitting} identify the field-induced splitting in their spectral function with the splitting in differential conductance and predict that the screening should cause $\DeltaK < \Delta$ at all fields and that \DeltaK\space decreases logarithmically with increasing \TKondo. Recent observations by Kogan \etal\cite{kogan04:SpinNKondoSplit} have confirmed the presence of the critical magnetic field \Bc. However the latter measurements, as well as those of Zumb\"{u}hl \etal\cite{zumbuhl04:cotunnelingspec}, show that $\DeltaK > \Delta$ at high fields, in disagreement with theory.  
	
	Here we extend the work of Kogan \etal\cite{kogan04:SpinNKondoSplit} by reporting measurements of \DeltaK~ as a function of the Kondo temperature \TKondo, as well as $B$. These measurements exploit the property of SETs that \TKondo~ varies continuously with gate voltage, in contrast with conventional Kondo systems, for which \TKondo~ is determined by chemistry. We find $\DeltaK > \Delta$ for high fields at the lowest Kondo temperatures, as observed previously. We also observe that \DeltaK~ decreases as $\ln(\TKondoZero/\TKondo)$ with increasing \TKondo, where \TKondoZero~ is a constant. This logarithmic decrease in \DeltaK~ with increasing \TKondo~ agrees qualitatively with theory. However, theory predicts that the decrease should be proportional to $\ZeemanE\ln(\TKondoZero/\TKondo)$, whereas we find the prefactor of the logarithm to be larger than predicted and independent of $B$. Our measurements also show that there is a critical magnetic field \Bc~ for splitting the Kondo peak, as predicted. However, we find that the predicted values of \Bc~ as a function of \TKondo~ appear to be too large.
	
	The SET we study is fabricated from a heterostructure consisting of an undoped GaAs buffer, followed by a $15$ nm layer of \AlxGaAs{0.3}{0.7} $\delta$-doped twice with a total of $10^{13}$ $\mbox{cm}^{-2}$ of Si, and finally a $5$ nm GaAs cap. The two-dimensional electron gas (2DEG) formed at the AlGaAs/GaAs interface has an electron density of $8.1\times10^{11}$ $\text{cm}^{-2}$ and a mobility of $10^{5}$ $\text{cm}^{2}$/Vs at $4.2$ K. Although the density of the 2DEG is high, magneto-transport measurements show that only a single subband is occupied. Electron-beam lithography is used to define the gate electrode pattern shown in the inset of \subfig{figure2}{b}. Applying a negative voltage to these electrodes depletes the 2DEG underneath them and forms an artificial atom of about $50$ electrons isolated by two tunnel barriers from the remaining 2DEG regions, the source and drain leads. The electrochemical potential of the dot, as well as the coupling between the dot and the leads, can be tuned by changing the voltages on the electrodes. The voltage on the gate electrode g is denoted \Vg. The SET is measured in a $75$ $\mu$W Oxford Instruments dilution refrigerator with an $8$ T magnet and a lowest electron temperature of about $100$ mK. To minimize orbital effects we align the 2DEG parallel to the magnetic field to within a few degrees. We measure the differential conductance \plaindidv\space utilizing standard lock-in techniques.

	To compare \DeltaK~ to $\Delta$ we need an accurate measurement of the Zeeman splitting, and hence $|g|$, in our SET. Such measurements were reported for this SET in reference 11 and the results are summarized here. We measured the Zeeman splitting in this SET using the traditional method of electron addition spectroscopy and the new, more precise method of inelastic spin-flip cotunneling. These measurements showed that $\Delta$ was linear with field up to $7.8$ T and the slope was given by $|g|= 0.16$. We also used inelastic cotunneling to measure the Zeeman splitting in an identical device in a different dilution refrigerator with a lower electron temperature and a larger magnet. In this setup, the device was aligned parallel to the magnetic field to better than $0.5$ degree. Measurements of this sample showed that $\Delta$ was linear with field up to a magnetic field of $14$ T and that $|g|$ was consistent with the value measured in the other device. We attribute the small $g$-factor to the penetration of the electron wavefunction into the AlGaAs. We believe the large electric fields produced by the relatively high electron density enhances this effect. Measurements of the Kondo splitting in a magnetic field in both devices showed that \DeltaK\space increased linearly with field with a slope given by $|g|$, however in both devices $\DeltaK > \Delta$ by about $10$ $\mu$eV. 
	
\begin{figure}

\begin{center}
\includegraphics[width=7.0cm, keepaspectratio=true]{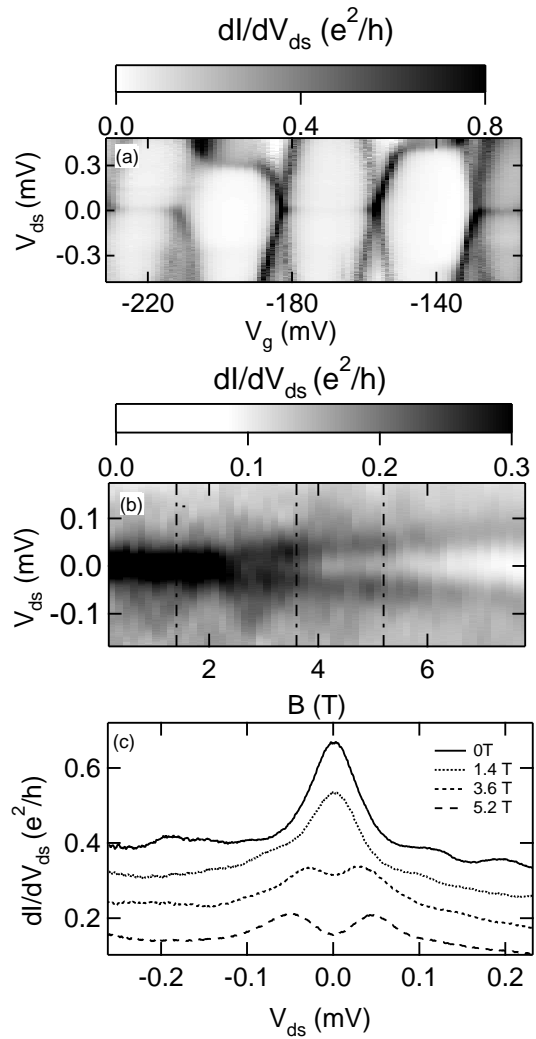}
\end{center}

	\caption{(a) Examples of Coulomb blockade diamonds when the artificial atom is strongly coupled to the leads. The Kondo effect shows up as sharp resonances near $\Vds= 0$ in the odd valleys. The inelastic cotunneling thresholds in the even valleys indicate the energies of the excited states. These data are at different electrode voltages and during a different cool-down of the device from those in the rest of the paper. (b) Evolution of a Kondo peak as a function of magnetic field. Note that the splitting does not begin until $B = 2.4$ T. (c) \plaindidv~ vs \Vds~ for $B= 0$ T and for other values of $B$ marked by the dash-dot lines in (b). For $B< 2.4$ T there is only one Kondo peak; above the threshold it splits. The traces have been offset by $0.08$ $e^{2}/h$ for clarity.    
	}
	\label{fig:figure1}
\end{figure}

 	We observe the Kondo effect by tuning the gate voltages so that the artificial atom is strongly coupled to the leads. \startsubfig{figure1}{a} shows an example of a plot of \plaindidv\space as a function of \Vds\space and \Vg\space for this device. Enhancements in the conductance at $\Vds= 0$ in the middle of every second Coulomb blockade valley are clearly visible. We infer that the valleys with Kondo peaks are those with an odd number of electrons. A sharp peak in \plaindidv\space at $\Vds= 0$, for \Vg~ between the Coulomb charging peaks where resonant tunneling is prohibited by energy and charge conservation, is the hallmark of the Kondo effect. The gate lithography makes this device very small, resulting in typical excited state energies on the order $300-400$ $\mu$eV. The large excited state energies can be seen explicitly in \subfig{figure1}{a} by the location of the orbital inelastic cotunneling thresholds in the even valleys. It has been demonstrated in both AlGaAs/GaAs and carbon nanotube SETs that a perpendicular magnetic field\cite{zumbuhl04:cotunnelingspec, sasaki2000:verticalST, vdwiel2002:lateralST, nygard2000:nanotube_ST} or gate voltage\cite{kogan2003:ST_zeroB, kyriakidis2002:voltage-tunableST, fuhrer2003:ringST} can reduce the energy of the excited state and lead to higher spin Kondo states. The large excitation energy in our device makes this transition unlikely, especially since our magnetic field is parallel to the 2DEG, and we see no evidence for this transition in our data. For the remainder of this paper we focus on studying spin-1/2 Kondo physics. 
 	
 	An example of a peak at $B= 0$ T is shown in \subfig{figure1}{c}. Applying a magnetic field causes the Kondo peak to split into two peaks above and below $\Vds=0$. This is shown in \subfig{figure1}{b}, while \subfig{figure1}{c} shows \plaindidv\space vs \Vds\space at various magnetic fields. To study the splitting of a Kondo peak in \plaindidv\space as a function of \TKondo, we take advantage of the fact that \TKondo\space varies across a Coulomb valley. The dependence of \TKondo\space on the energy $\epsilon_{0}$ of an unpaired electron in the artificial atom referenced to the Fermi level is predicted to be\cite{goldhaber98:PRLKondo, haldane78:ScalingTheory}
\begin{equation}       
\TKondo = \frac{\sqrt{\Gamma U}}{2\kboltz} \exp[ \pi\epsilon_{0}(\epsilon_{0}+U)/\Gamma U ].
\label{eq:OriginalHaldaneEqn}
\end{equation}		
Here $\Gamma$ is the full width at half-maximum of the Coulomb charging peaks and $U$ is the charging energy necessary to add an electron to the artificial atom. Goldhaber-Gordon \etal\cite{goldhaber98:PRLKondo} have shown that their measurements of \TKondo~ are fit well by this equation. Expressing \TKondo~ in terms of its value in the middle of the Coulomb valley    

\begin{equation}       
\TKondo = \TKondoZero \exp[ \pi(\Delta\epsilon)^{2}/\Gamma U ].
\label{eq:HaldaneEqn}
\end{equation}

In this equation, \TKondoZero~ is the Kondo temperature in the middle of the Coulomb valley, and $\Delta\epsilon$ is the difference in the energy from its value in the middle of the valley. $\Delta\epsilon$ is related to the gate voltage by $\Delta\epsilon = \alpha_{g}e(\Vg - \VgZero)$ where $\alpha_{g}$ is the ratio of the gate capacitance to the total capacitance and \VgZero~ is the value of \Vg~ in the middle of the valley. We can thus re-write the argument of the exponential in \refeqn{HaldaneEqn}~ as $\chi(\Vg - \VgZero)^{2}$ where $\chi= \pi\alpha_{g}^{2}e^{2}/\Gamma U$. The Kondo temperature is a minimum in the middle of the Coulomb valley and increases exponentially close to the Coulomb blockade peaks. Thus, by varying \Vg, we can study the variation in \DeltaK~ as a function of \TKondo. If the Kondo temperature gives a logarithmic correction to the splitting as predicted,\cite{moore00:KondoSplitting} then we expect \DeltaK~ to vary quadratically in gate voltage. 

	To make quantitative comparisons with theory we need to determine \TKondo~ for a range of \Vg. To determine \TKondo~ at a fixed gate voltage we follow Goldhaber-Gordon \etal\cite{goldhaber98:PRLKondo} and measure the differential conductance at $\Vds=0$ (denoted $G$) as a function of temperature. We then fit these data to an empirical form that gives a good approximation to numerical renormalization group results\cite{costi94:TransportCoef} to obtain the Kondo temperature \TKondo. The empirical form is given by
\begin{equation}       
	G(\mbox{T}) = G_{0}[1+(2^{1/s}-1)(\mbox{T}/\TKondo)^{2}]^{-s}
\label{eq:EmpiricalForm}
\end{equation}	
where $s= 0.22$ for the spin-1/2 Kondo effect. This fit is accurate only if we have data for temperatures as low as $\mbox{T}\sim 0.1\TKondo$. The lowest temperature we can achieve is about $100$ mK, so we can reliably measure Kondo temperatures larger than approximately $1$ K. The Kondo temperature in the middle of the Coulomb blockade valley is significantly lower than this, so we cannot measure it directly. However, we can measure \TKondo~ near the Coulomb blockade peaks where the Kondo temperature is much higher. Fitting the latter data for \TKondo~ as a function of \Vg~ to \refeqn{HaldaneEqn}~ we can determine \TKondoZero~ and $\chi$. 

\begin{figure}[!]
\setlength{\unitlength}{1cm}
\begin{center}
\begin{picture}(8,8.75)(0,0)
\put(0,0){\includegraphics[width=8cm, keepaspectratio=true]{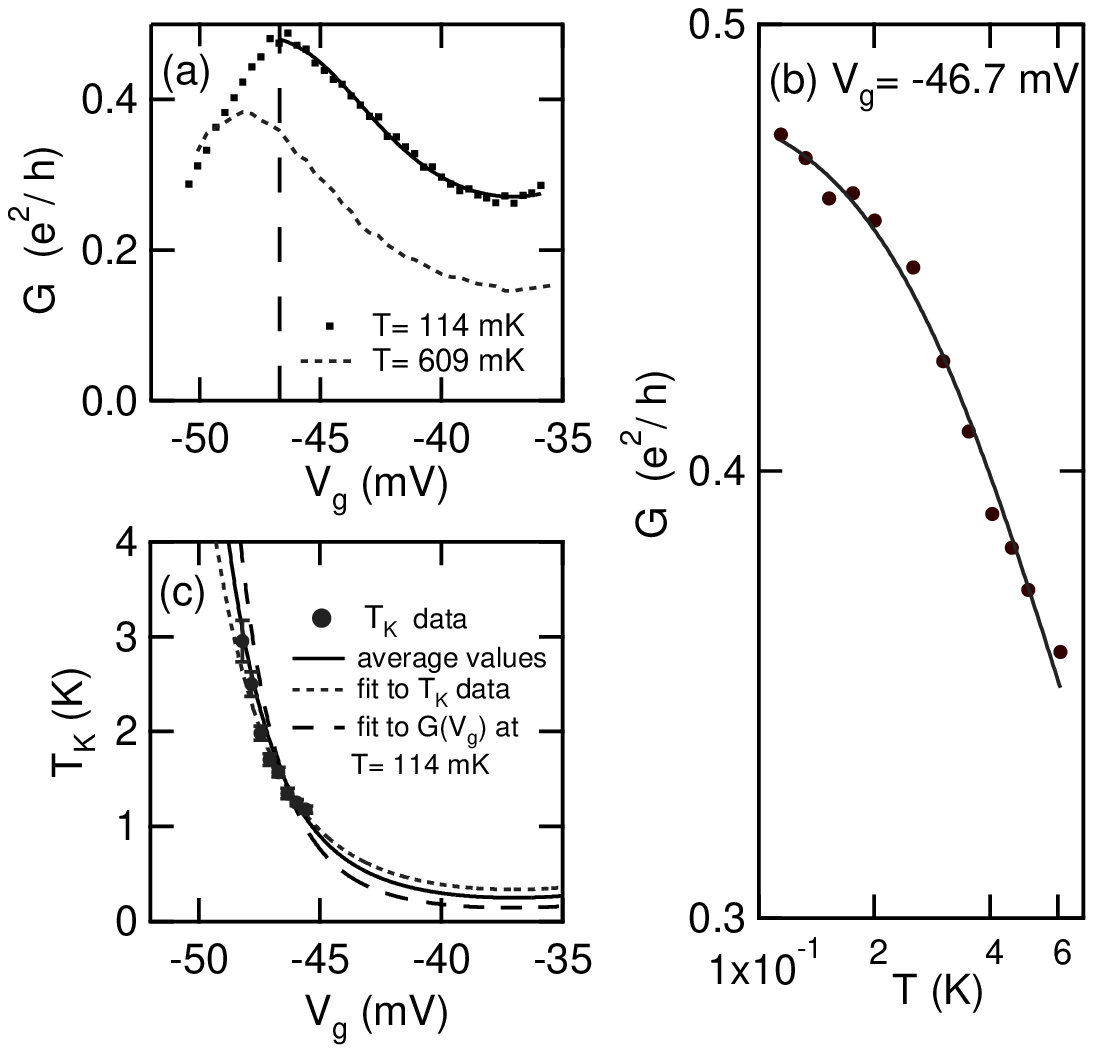}}
\put(5.7,1.2){\includegraphics[width=2cm, keepaspectratio=true]{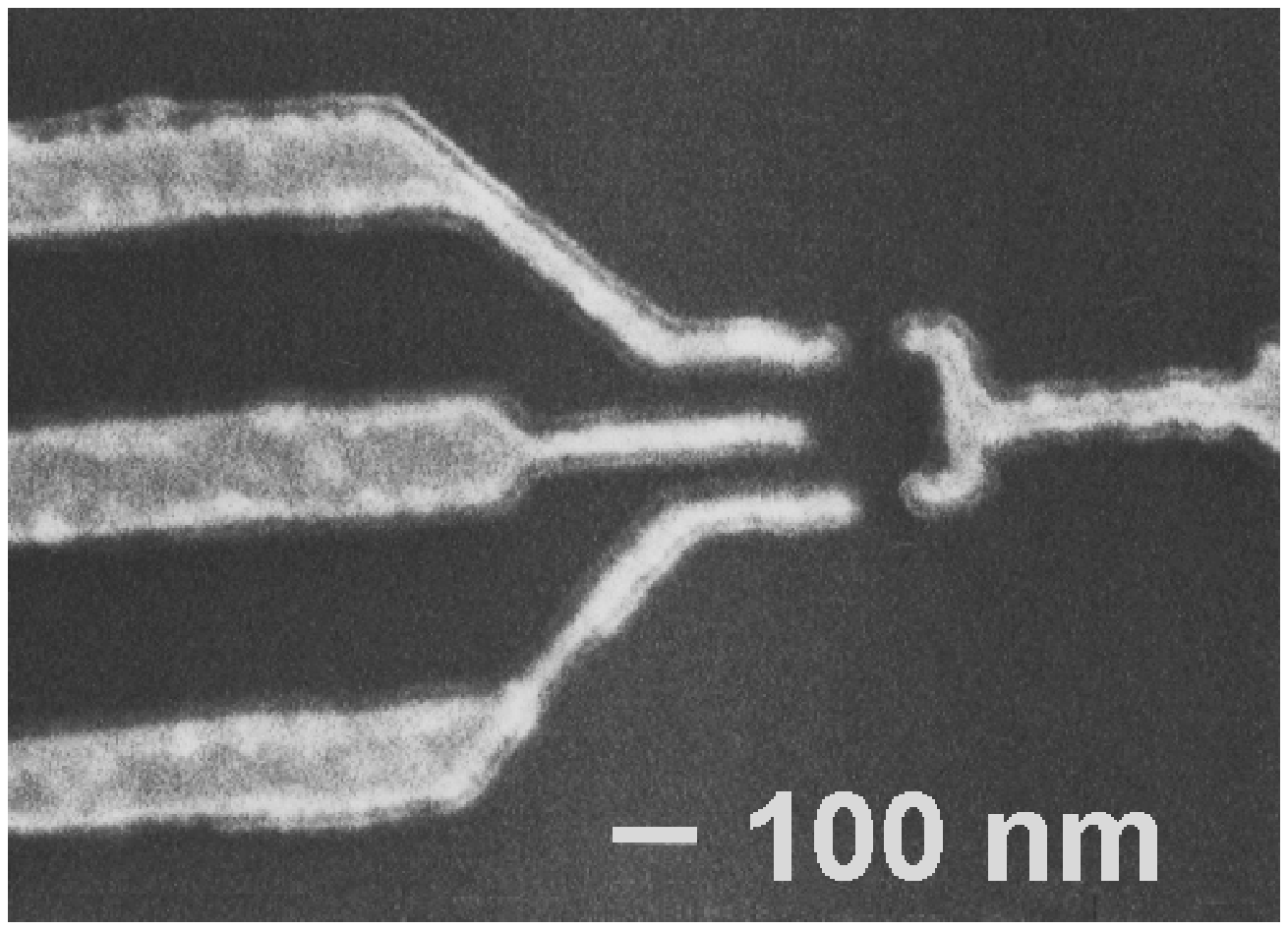}}
\put(5.3,1.9){\makebox(0,0){g}}
\put(5.45,1.9){\line(1,0){0.4}}
\end{picture}
\end{center}

	\caption{(a) \plaindidv~ vs \Vg~ at $\Vds= 0$ (denoted $G$) for a Coulomb charging peak and half a Coulomb valley at two different temperatures. The solid line is a fit to Equations \ref{eq:HaldaneEqn} and \ref{eq:EmpiricalForm} and is discussed in the text. (b) $G$ as a function of temperature for the value of \Vg~ marked by the dashed vertical line in (a). Fitting these data to \refeqn{EmpiricalForm}~ gives $\TKondo= 1.57 \pm 0.05$ K. The inset shows an electron micrograph of an SET similar to the one we have studied. (c) The dots show \TKondo~ determined as in (b) as a function of \Vg. The dotted line is the result of fitting these data to \refeqn{HaldaneEqn}. The dashed line shows \TKondo~ as a function of \Vg~ using the results of the fit shown in (a). The solid black line shows \TKondo~ for the parameter values $\chi = 0.020$ $\mbox{(mV)}^{-2}$ and $\TKondoZero= 250$ mK.
	}
	\label{fig:figure2}
\end{figure}

	Figure \ref{fig:figure2} illustrates this procedure. \startsubfig{figure2}{a} shows a Coulomb charging peak and half the Coulomb valley at two different temperatures. As the temperature is raised, the suppression of the Kondo effect removes the energy renormalization, causing the peak to move toward the bare resonance.\cite{goldhaber98:PRLKondo, wingreen94:AndersonOutofEquil} We measure $G$ as a function of temperature at several values of \Vg, one of which is shown in \subfig{figure2}{b}, and fit these data to the empirical form given by \refeqn{EmpiricalForm}~ to determine \TKondo. Finally, the data points in \subfig{figure2}{c} shows \TKondo~ as a function of \Vg~ for the range in which it can be determined with confidence. Fitting these data to \refeqn{HaldaneEqn}, shown by the dotted line in \subfig{figure2}{c}, we find $\TKondoZero= 340$ mK and $\chi = 0.016$ $\mbox{(mV)}^{-2}$. 

  Another way of determining \TKondo~ is to take advantage of the fact that $G_{0}$ is independent of gate voltage between the bare Coulomb resonances\cite{goldhaber98:PRLKondo}. Combining Equations \ref{eq:HaldaneEqn} and \ref{eq:EmpiricalForm} gives an equation for $G(\Vg)$ at a fixed temperature with fit parameters $\chi$,\TKondoZero, and $G_{0}$. We use this equation to fit the $G(\Vg)$ data at $\mbox{T}= 114$ mK, shown in \subfig{figure2}{a}. The fit is shown as the solid line through the data and gives $\TKondoZero= 145$ mK and $\chi = 0.026$ $\mbox{(mV)}^{-2}$. A plot of \TKondo~ vs \Vg~ for these fit parameters is shown by the dashed line in \subfig{figure2}{c}. From these two methods, we arrive at the values $\chi = 0.020 \pm 0.006$ $\mbox{(mV)}^{-2}$ and $\TKondoZero= 250 \pm 100$ mK. A plot of \TKondo~ vs \Vg~ determined from these parameters is shown as the solid line in \subfig{figure2}{c}. We can use our value of $\chi$ to estimate $\Gamma$. From nearby Coulomb charging diamonds, we estimate $U= 1.2$ mV and $\alpha_{g}= 0.05$. Combining this with our value of $\chi= 0.020$ $\mbox{(mV)}^{-2}$ we find that $\Gamma\approx 330$ $\mu$eV. Our values are similar to those obtained by Goldhaber-Gordon \etal,\cite{goldhaber98:PRLKondo} who found $U= 1.9 \pm 0.05$ meV, $\alpha_{g}= 0.069 \pm 0.0015$, and $\Gamma= 280 \pm 10$ $\mu$eV, and are consistent with having a somewhat larger artificial atom with stronger coupling to the leads. As expected for this stronger coupling, \TKondoZero~ in our SET is larger than that of Goldhaber-Gordon \etal\cite{goldhaber98:PRLKondo}

\begin{figure}[!]
\begin{center}
\includegraphics[width=6cm, keepaspectratio=true]{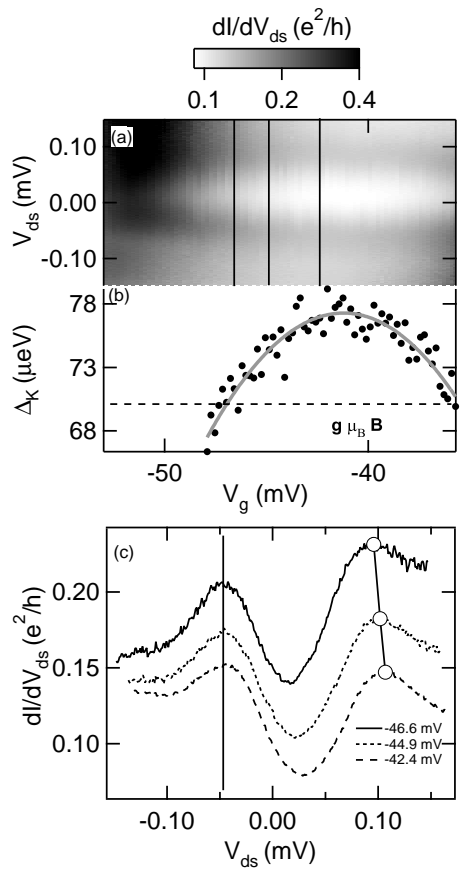}
\end{center}

	\caption{(a) A Kondo feature in a $7.5$ T magnetic field. The Kondo peak at zero bias has clearly split into two peaks. One Coulomb charging peak is visible at $\Vg= -52$ mV; the other peak is not visible. (b) \DeltaK~ as a function of \Vg. The splitting is a maximum in the center of the valley, where the Kondo temperature is lowest. The gray line shows the result of fitting the data to a parabola. The dashed horizontal line indicates the Zeeman energy \ZeemanE. Note that the splitting in the middle of the valley exceeds \ZeemanE. (c) \plaindidv~ vs \Vds~ for the values of \Vg~ marked by the lines in (a). The traces have been shifted horizontally so that the peaks at negative \Vds~ are aligned. The circles mark the positions of the peaks at positive \Vds, showing that the splitting between the peaks changes with gate voltage.   
	}
	\label{fig:figure3}
\end{figure}
   
	To study the dependence of \DeltaK~ on \TKondo~ we measure \plaindidv~ as a function of \Vds~ and \Vg~ at various magnetic fields. An example of such a measurement is shown in \subfig{figure3}{a}, while \subfig{figure3}{c} shows \plaindidv~ vs \Vds~ traces at the values of \Vg~ marked in \subfig{figure3}{a}. We determine the position of a peak by fitting the data points in the neighborhood of the peak to a parabola. The center of the parabola determines the peak position. At a given \Vg, \DeltaK~ is half the separation between the positions of the positive and negative peaks. The values of \DeltaK~ as a function of \Vg~ are shown in \subfig{figure3}{b}. These data are fit very well by a parabola, which is shown in the figure. As discussed above, a parabola is what one would expect from a correction to \DeltaK~ that is logarithmic in the Kondo temperature.
	
\begin{figure}[!]

\begin{center}
\includegraphics[width=6cm, keepaspectratio=true]{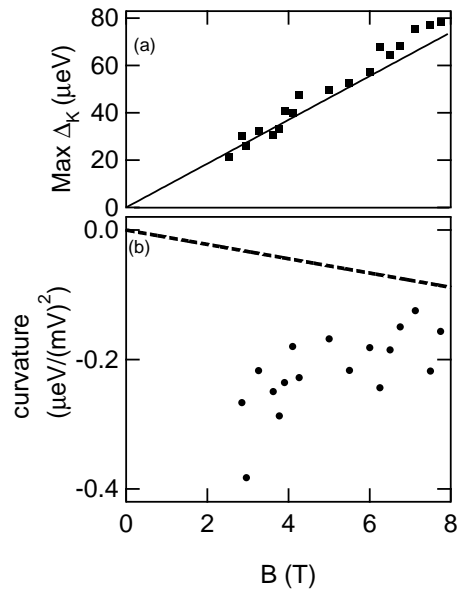}
\end{center}

	\caption{(a) Splitting of the Kondo peak in the center of the Coulomb valley as a function of magnetic field. The solid line shows $\Delta=\ZeemanE$ for $|g|= 0.16$ determined by inelastic cotunneling. Above $\sim 4$ T we find that $\DeltaK > \ZeemanE$. Below $B= 2.4$ T no splitting is observed. (b) Curvature of \DeltaK~ vs \Vg~ parabolas like that in \subfig{figure3}{b}. The prediction of Moore and Wen\cite{moore00:KondoSplitting} (dashed line) is that the curvature should decrease with field.
	}
	\label{fig:figure4}
\end{figure}	
	
	One important feature of these data is that for $B > 4$ T, the maxima of the the \DeltaK~ vs \Vg~ curves lie above $\Delta= \ZeemanE$, where the value of $\Delta$ is determined precisely using inelastic cotunneling.\cite{kogan04:SpinNKondoSplit} By fitting $\DeltaK(\Vg)$ to a parabola, we extract the maximum value of \DeltaK, which is plotted as a function of magnetic field in \startsubfig{figure4}{a}. The solid line shows the value of $\Delta$ from cotunneling measurements. These data show that for large magnetic fields, $\DeltaK > \Delta$ by about 10 $\mu$eV as observed previously.\cite{kogan04:SpinNKondoSplit, zumbuhl04:cotunnelingspec}

	Another important feature of these data is the curvature of the parabolas, which is plotted versus $B$ in \subfig{figure4}{b}. Moore and Wen\cite{moore00:KondoSplitting} predict that for $1 <\Delta/\kboltz \TKondo< 100$, $\DeltaK / \Delta =b\ln(\Delta/\kboltz\TKondo)+constant$ where $b$ is a constant of proportionality and is approximately $0.06$. From \refeqn{HaldaneEqn}, \TKondo~ varies exponentially with gate voltage so that at fixed magnetic field the prediction is $\DeltaK = \DeltaKZero - b(\ZeemanE)\chi(\Vg - \VgZero)^{2}$, where \DeltaKZero~ gives the field splitting at the center of the valley. This is the equation of a parabola in \Vg~ whose curvature becomes more negative with increasing magnetic field. This prediction is plotted in  \subfig{figure4}{b} for $\chi = 0.020$ $\mbox{(mV)}^{-2}$ and clearly does not agree with the data. The magnitude of the curvature is much larger than predicted by theory, except at the highest fields, and does not grow with increasing magnetic field. The constant curvature of the data indicates that the correction to the Kondo splitting is given by $c\ln(\TKondoZero/\TKondo)$, where $c$ is a constant; by comparing to the data we find $c = 12$ $\mu$eV. 

\begin{figure}[!]

\begin{center}
\includegraphics[width=8cm, keepaspectratio=true]{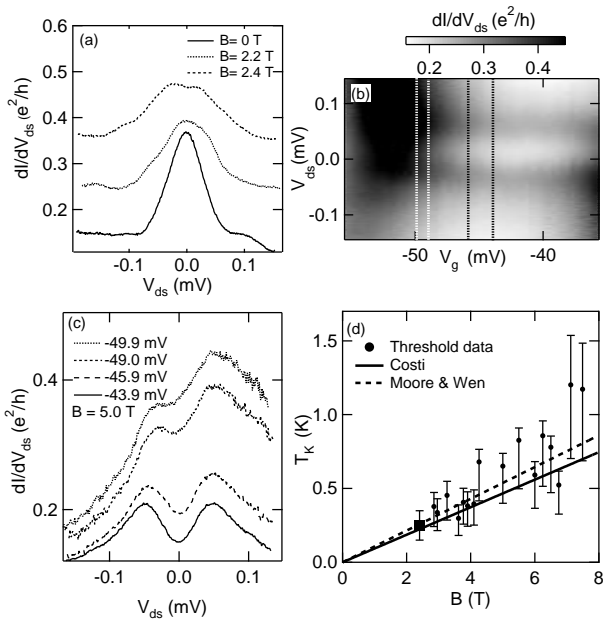}
\end{center}

	\caption{(a) Drain-source sweeps in the middle of a Kondo valley at different magnetic fields. The onset of splitting is clear at $B= 2.4$ T but not visible at $2.2$ T. The curves have been offset by $0.11$ $e^{2}/h$ for clarity. (b) Kondo valley with a $5$ T field applied. The lines indicate the positions of the \plaindidv~ vs \Vds~ traces shown in (c). (c) \plaindidv~ vs \Vds~ at various gate voltages. In the center of the valley where \TKondo~ is a minimum the splitting is clear and symmetric. Closer to the charging peak the Kondo temperature is higher and the splitting is less pronounced, but still visible. (d) The dots mark the maximum \TKondo~ at which splitting is observed as a function of field. These data are extracted from data like those in (b) and (c) by measuring the gate voltage at which the splitting is barely visible. The conversion of this gate voltage to a Kondo temperature is explained in the text. The square data point at $B= 2.4 T$ marks the measurement of \Bc~ from the data in (a). Predictions for \Bc~ of Costi\cite{costi00:KondoSplitting} (solid line) and Moore and Wen\cite{moore00:KondoSplitting} (dashed line) are shown. 
	}
	\label{fig:figure5}
\end{figure}

	We can also use these data to test predictions about the threshold field \Bc~ necessary to split the Kondo peak. The existence of a threshold field is illustrated in \subfig{figure1}{b}. The peak does not start to split until $B$ is greater than about $2.4$ T. \startsubfig{figure5}{a} shows this threshold in more detail: at $B= 2.0$ T the peak has not yet split, but by $2.4$ T, the splitting is clearly evident. We can also extract information about \Bc~ from data like those in \allfig{figure5}. For a given magnetic field there is some Kondo temperature \TKondoC~ for which this field is the critical field. For \TKondo~ below \TKondoC, the Kondo peak is split and for \TKondo~ greater than \TKondoC~ the peak is not split. This is illustrated by the data in \subfig{figure5}{b} and in \subfig{figure5}{c}, which shows some individual traces from \subfig{figure5}{b}. In the middle of the valley, where the Kondo temperature is lowest, the peak is clearly split. This splitting disappears toward the Coulomb charging peak where the Kondo temperature is higher. Because of the background from the Coulomb charging peak, it is difficult to reliably locate \TKondoC. Instead, we identify a range of Kondo temperatures over which we are confident that the peak is split: this puts a lower bound on \TKondoC. 
	
	To extract this bound, we locate the most negative gate voltage at which we still confidently observe the Kondo splitting. For the data in \subfig{figure5}{b} and (c), this is at $\Vg= -49.9$ mV. To convert this to a Kondo temperature we use \refeqn{HaldaneEqn}. To find the center of the valley \VgZero, we use the location of the maximum of the fit of \DeltaK(\Vg) to a parabola. We take into account the uncertainty in the determination of \VgZero~ to make the most conservative estimate of the lower bound. Knowing \VgZero~ we can convert \Vg~ into a Kondo temperature using \refeqn{HaldaneEqn}. The lower bounds on \TKondoC~ found in this way using $\chi = 0.020$ $\mbox{(mV)}^{-2}$ and $\TKondoZero= 250$ mK are shown as data points in \subfig{figure5}{d}. The error bars show the Kondo temperatures extracted using the other two sets of $\chi$ and \TKondoZero~ shown in \subfig{figure2}{c} and taking into account errors on these values. At Kondo temperatures below these data points, the Kondo peak is split. 
	
	We can use our data to check predictions about \TKondoC. Costi\cite{costi00:KondoSplitting} predicts that $\Bc = 0.5 \THWHM$, where \THWHM~ is the half width at half-maximum of the Kondo resonance in the spectral function at $T=0$ and is related\cite{costi00:KondoSplitting,costi04:PrivateCom} to \TKondo~ by $\TKondo = 0.433 \THWHM$. This prediction is shown as the solid line in \subfig{figure5}{d}: Costi predicts that below this line the Kondo peak should be split and above it the Kondo peak should not be split. However, this line lies below the lower bound set by some of our data points suggesting that at a given \TKondo, \Bc~ is smaller than predicted by theory. The dotted line in \subfig{figure5}{d} is based on the work of Moore and Wen, who predict\cite{moore00:KondoSplitting} the width and splitting of the spin-up component of the spectral function as a function of $B/\TKondo$. To model the Kondo peak arising from this spectral function we sum two lorentzians\cite{moore04:PrivateCom} with the width and splitting given by the predictions of Moore and Wen. We find that the splitting  appears at $\Bc = \TKondo$. This prediction is plotted as a dashed line in \subfig{figure5}{d} and lies below several data points at higher magnetic fields. 
	
	The observation that $\DeltaK > \Delta$ by Kogan \etal\cite{kogan04:SpinNKondoSplit} and Zumb\"{u}hl \etal\cite{zumbuhl04:cotunnelingspec} demonstrates that we do not yet have a full theoretical understanding of the non-equilibrium Kondo effect. Our measurements of the Kondo temperature dependence of \DeltaK~ re-enforce this conclusion. While theory qualitatively describes the behavior of \DeltaK~ with \TKondo, it does not provide an accurate quantitative description. We find that \DeltaK~ decreases logarithmically with increasing \TKondo, but the prefactor of the logarithm is larger than expected at low fields and is independent of $B$. There is clear evidence of a critical field \Bc~ for the splitting of the Kondo peak, again in qualitative agreement with theory. However, our data suggests that \Bc~ is smaller than predicted by theory. Our results provide additional challenges to the theory of the non-equilibrium Kondo effect.        

	We thank D. Goldhaber-Gordon and D. Mahalu for designing and fabricating the SET devices used in this 
work and H. Shtrikman for growing the GaAs/AlGaAs heterostructures. We are grateful to J. Moore, T. Costi, L. Levitov, X.-G. Wen, W. Hofstetter, C. Marcus, D. Zum\"{u}hl, and J. Folk for discussions and to C. Cross, G. Granger, and K. MacLean for experimental help. This work was supported by the US Army Research Office under Contract DAAD19-01-1-0637, by the National Science Foundation under Grant No.~DMR-0102153, and in part by the NSEC Program of the National Science Foundation under Award Number DMR-0117795 and the MRSEC Program of the National Science Foundation under award number DMR 02-13282.	 

% Bibliography:	
%\bibliography{samasha_STKondoref}

\end{document}